\documentclass[prb,floatfix,twocolumn,showpacs,preprintnumbers,amsmath,amssymb]{revtex4}
\usepackage[dvips]{graphics}
\usepackage{amssymb}
\usepackage{psfig}

\begin{document}

\title{Resonant amplification of the Andreev process in ballistic Josephson junctions}

\author{Ivana Petkovi\'{c}$^{1,2}$, Nikolai
M.~Chtchelkatchev$^{3,4}$, and Zoran Radovi\'{c}$^1$}
\address{\small$^1$ Department of Physics,University of Belgrade, P.O. Box 368, 11001 Belgrade, Serbia and Montenegro \\
\small$^2$ Laboratoire de Physique des Solides, Batiment 510, 91405
Orsay, France \\ \small$^3$ L.D.\ Landau Institute for Theoretical
Physics, RAS, 117940 Moscow, Russia \\
\small$^4$ Institute for High Pressure Physics, RAS, 142092 Troitsk
, Russia}


\begin{abstract}
We study the Josephson effect in ballistic SINIS and SIFIS
double-barrier junctions, consisting of superconductors (S), a clean
normal (N) or ferromagnetic (F) metal, and insulating interfaces
(I). For short SINIS double-tunnel junctions with one channel open
for quasiparticle propagation, the critical Josephson current as a
function of the junction width shows sharp peaks because of a
resonant amplification of the Andreev process: when the quasibound
states of the normal interlayer enter the superconducting gap the
Andreev bound states are lowered down to the Fermi level. For
corresponding SIFIS junctions the quasibound states are spin-split;
they amplify the supercurrent less efficiently, and trigger
transitions between $0$ and $\pi$ states of the junction. In
contrast to SINIS junctions, a narrow dip related to the $0$ --
$\pi$  transition opens up exactly at the peak due to a compensation
of partial currents flowing in opposite directions. With an
increased barrier transparency these features are gradually lost,
due to the broadening and overlap of quasibound states.
Temperature-induced transitions both from $0$ to $\pi$ and from
$\pi$ to $0$ states are studied by computing the phase diagram (with
temperature and junction width as the variables) for different
interfacial transparencies varying from metallic to the tunnel
limit.
\end{abstract}


\maketitle

\section{Introduction}

Recently, many interesting phenomena were investigated in
superconductor (S) - ferromagnet (F) Josephson
junctions.\cite{buzdin_revmod,golubov_revmod} One of the most
interesting effects is the so-called $\pi$ state of SFS junctions,
in which the ground state is characterized by an intrinsic phase
difference $\phi = \pi$ between two superconductors.
\cite{bulaevskii,bulaevskii_buzdin,kiza,ryazanov,kontos,bauer,jiang,obi}
The $\pi$ junctions could also have important applications, such as
for quantum qubits. \cite{taro} The Josephson effect in
double-barrier SIFIS junctions, with a clean metallic ferromagnet
between two insulating interfaces (I), has been studied by solving
the scattering problem based on the Bogoliubov - de Gennes equation;
it was found that spin-split quasiparticle transmission resonances
contribute significantly to the Andreev process.\cite{greek,cayssol}
A comprehensive numerical study of clean layered S/F
heterostructures has also been given recently.\cite{valls}

For SINIS structures (with a clean non ferromagnetic normal metal
N), sharp peaks occur in the critical Josephson current $I_C$ as a
function of the junction width $d$ for each transverse channel. This
is the result of a resonant amplification of the Andreev process
when the quasibound states in the INI interlayer approach the Fermi
level. This effect is the superconducting analog of the Bohm
resonant tunneling,\cite{bohm} previously studied in
Refs.~\cite{Beenakker,beenakker_van_houten,melsen_beenakker,ketterson}
For corresponding SIFIS junctions, the resonant nature of $0$ --
$\pi$ transitions comes into play in the tunnel limit. Triggering of
the $0$ -- $\pi$  transitions by resonant amplification, as well as
the coexistence of stable and metastable $0$ and $\pi$ states in the
vicinity of the transition,\cite{2001,sellier} can be fully
understood. In contrast to SINIS junctions where the critical
current reaches a peak value when the Andreev bound states cross the
Fermi level, here a narrow dip related to the $0$ -- $\pi$
transition opens up exactly at the peak due to the compensation of
partial currents flowing in opposite directions. We emphasize that
this is not the case for a higher interface transparency due to the
broadening and overlap of quasibound states, nor for planar
junctions.\cite{greek} Experimentally, SIFIS nanostructures could be
realized, for example, in gated heterostructures,\cite{Takayanagi}
ferromagnetic nanoparticles, \cite{sophie} or in the break
junctions\cite{Scheer} in external magnetic field producing the
Zeeman splitting of Andreev levels.\cite{Yip}

It has become a paradigm that in the $\pi$ state of SFiS junctions,
with a thin barrier of insulating ferromagnet (Fi), there is net
spin polarization of current-carrying Andreev
levels.\cite{tanaka,Nikolai_SFSpi,barash} However this ``rule'' is
not quite general. In this paper we show that in low-transparent
SIFIS nanostructures with one open transverse channel the situation
can be the opposite. We can generalize the previous rule: the $0$ --
$\pi$ transition is accompanied by the switching of the net spin
polarization of current-carrying Andreev levels. The critical
current oscillates as a function of the junction width, and the
resonant amplification occurs when quasibound states of the
ferromagnet between two insulating layers enter the gap. The $0$ --
$\pi$ transitions take place when Andreev levels intersect at the
Fermi energy, being accompanied by a switching of their spin
polarization. Experiments show that the $\pi$ state of a dirty SFS
junction can correspond to lower temperatures than the $0$
state.\cite{ryazanov} The theory of ballistic one-barrier SFS
junctions predicts the opposite.\cite{Nikolai_SFSpi,nick_jetp} We
demonstrate that in the phase diagram of double-barrier ballistic
SIFIS junctions the $\pi$ state can correspond to lower temperatures
than the $0$ state, depending on parameters.

\section{Josephson Current}

\begin{figure}[h]
\centerline{\hbox{
  \psfig{figure=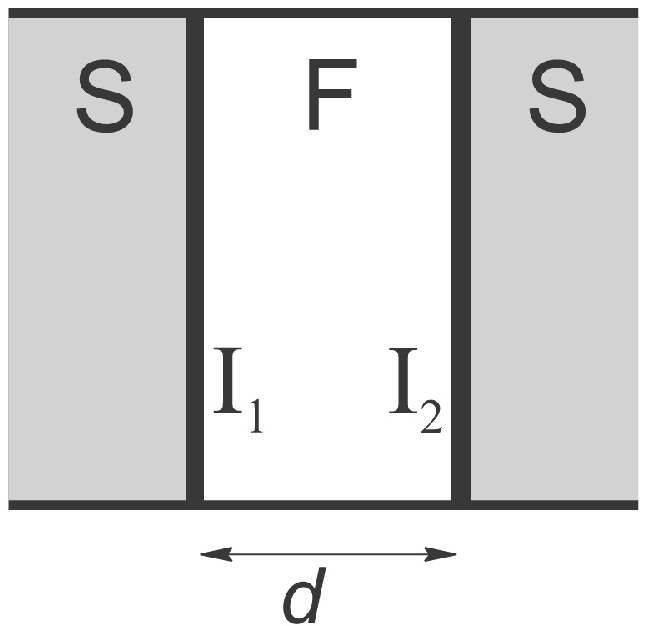,height=55mm,width=65mm,angle=0}
}}  \caption{} \label{PetkovicFig1}
\end{figure}

We consider a ballistic double-barrier Josephson junction: a
normal-metal (N) or ferromagnetic (F) interlayer of thickness $d$ is
connected to superconductors (S) by insulating non-magnetic
interfaces (I$_1$, I$_2$), see Fig.~\ref{PetkovicFig1}. We assume
that both metals are clean and that the left and the right
superconductors are equal with the constant pair potential $\Delta
\exp(\pm i \phi/2)$, where $\phi$ is the macroscopic phase
difference across the junction. Within the Stoner model, the
ferromagnet is characterized by the exchange potential $h$, and the
insulating interfaces are modeled by $\delta$-function potential
barriers.\cite{greek,Nikolai_SFSpi}

The Josephson current can be expressed in terms of the free energy
of the junction as

\begin{equation}
I(\phi)=-\frac{2e}{\hbar}\partial_\phi F(\phi),
\end{equation}

\noindent where the phase-dependent part of the junction free
energy, $F(\phi)$, is given by

\begin{equation}
F(\phi)=-k_BT\sum_{n \sigma} \textrm{ln} \left[2
\textrm{cosh}\left(\frac{E_{n\sigma}(\phi)}{2k_BT}\right)\right].
\end{equation}


\noindent Here, $E_{n\sigma}(\phi)$ are the Andreev levels measured
with respect to the chemical potential, $n$ denotes a transverse
channel, and $\sigma=\pm 1$ is the spin quantum number.

Assuming the SIFIS junction shorter than the superconducting
coherence length, we can calculate the Josephson current using the
relation

\begin{equation}
I(\phi)=\frac{2e}{\hbar}\sum_{n \sigma} f_{n \sigma}(T) ~\partial
_\phi E_{n \sigma}(\phi),
\end{equation}

\noindent where $f_{n \sigma}(T)$ is the Fermi distribution, and
$E_{n\sigma}(\phi)$ are zeroes of the common denominator of the
scattering amplitudes $G$ , calculated from the Bogoliubov - de
Gennes equation and given by Eq. (17) in Ref. \cite{greek}

 In the following, we assume that only one channel ($n=1$) is open
in the short  superconducting junction. Without a ferromagnetic
region there are two spin degenerate Andreev levels with the same
absolute value of energy, but with the opposite sign that can be
parametrized by $\eta=\pm 1$. The ferromagnet spin-splits Andreev
levels, and there are four of them, $E_{\eta\sigma}(\phi)$. We will
express $E_{\eta\sigma}(\phi)$ explicitly in terms of the scattering
matrix.\cite{Nikolai_SFSpi,nick_jetp} For the IFI structure formed
by the ferromagnet and two insulating barriers we define $t_\pm$ and
$r_\pm$ to be the absolute values of the electron and hole
transmission and reflection amplitudes, $\hat t(\pm E, \sigma)$ and
$\hat r(\pm E,\sigma)$, where $E$ is the quasiparticle energy with
respect to the chemical potential. The corresponding phases are
denoted as $\chi_\pm^t$ and $\chi^r_\pm$. The Andreev levels can be
calculated as roots of the equation
\begin{equation}
\label{qc}
    \cos(\Phi-2\alpha)-\cos\gamma=0,
\end{equation}
\noindent where $\gamma$ is defined by 
\begin{equation}
\label{qc1}
    \cos\gamma=r_+r_-\cos\beta+t_+t_-\cos\phi.
\end{equation}

\noindent Here,
$\alpha=\arccos(E/\Delta)$,$~\Phi=\chi_+^t-\chi_-^t$, and
$\beta=(\chi_+^t-\chi_+^r)-(\chi_-^t-\chi_-^r)$. From Eq.~(\ref{qc})
the Andreev levels are

\begin{equation}
    E_{\eta\sigma}(\phi)=\Delta
    \cos\left(\frac{\Phi_0+\eta\gamma_0}2\right)
    {\rm sign}\left[\sin\left(\frac{\Phi_0+
    \eta\gamma_0}{2}\right)\right],
\end{equation}

\noindent where $\Phi_0=\Phi(E=0,\sigma)$ and
$\gamma_0=\gamma(E=0,\sigma)$. In the limiting case of a single
insulating barrier this formula reduces to the corresponding
equations found in Refs.~[\cite{Nikolai_SFSpi,barash,nick_jetp}].
Using Eq.~(\ref{qc1}) in the form
$\partial_\varphi\gamma=t_+t_-\sin\phi/\sin\gamma$, we finally
obtain

\begin{equation}\label{current}
    I(\phi)=\sum_{\sigma,\eta=\pm 1}
    \frac{e\Delta}{2 \hbar} \frac{t_+t_-\sin\phi}{\sin\gamma_0}
    \sin\left(\frac{\gamma_0+\eta\Phi_0}2\right)
    \tanh\frac{E_{\eta\sigma}(\phi)}{2 k_B T},
\end{equation}

\noindent which is a generalization of Eq.~(3) in
Ref.~[\cite{Nikolai_SFSpi}]. The transmission amplitudes of the IFI
structure can be expressed through the transmission and reflection
amplitudes of the insulating barriers, $\hat t_{I_{1,2}}$, $\hat
r_{I_{1,2}}$, and the momentum of an electron in F,
$q_\sigma(E)=\sqrt{2m({E_F}^{(F)}+E+\sigma h)}$, and in S,
$k=\sqrt{2m({E_F}^{(S)}+\Omega)}$, where
$\Omega=\sqrt{E^2-\Delta^2}$. Namely,

\begin{eqnarray}
\hat t(E,\sigma)&=&\frac{\hat t_{I_{1}}\hat
t_{I_{2}}e^{i(q_\sigma-k)d}}{1-\hat r'_{I_1} \hat r_{I_2}e^{2i
q_\sigma d}},
\nonumber\\
\hat r(E,\sigma)&=&r_{I_1}e^{-ikd}+\frac{(\hat
t_{I_{1}}e^{i(q_\sigma-k)d/2})^2r_{I_2}e^{i q_\sigma d}}{1-\hat
r'_{I_1}\hat r_{I_2}e^{2i q_\sigma d}}.
\end{eqnarray}

\noindent Note that the expression for the transmission amplitude
has the same form as the one for the Fabry-Perot interferometer.
Here, the scattering amplitudes of the insulating barriers are

\begin{eqnarray}
   \hat t_{I_{1,2}}=\frac{2\sqrt{k q_\sigma}}{k+q_\sigma+2iZ_{1,2}k_F^{(S)}},\nonumber\\
   \hat r_{I_{1,2}}=\frac{k-q_\sigma-2iZ_{1,2}k_F^{(S)}}{k+q_\sigma+2iZ_{1,2}k_F^{(S)}},
\end{eqnarray}

\noindent where $k_F^{(S)}=\sqrt{2m{E_F}^{(S)}}$, the parameters
$Z_{1,2}$ determine transmission probabilities of insulating
barriers, and $r'=-r^*t/t^*$ ($^*$ denotes the complex conjugation).
In the tunnel limit, $|\hat{t}_{I_{1,2}}|\to 0$, the positions of
the bound states are simply given by the conditions $d
q_\sigma(+E)=n_1 \pi$ and $d q_\sigma(-E)=n_2 \pi$, where $n_1$ and
$n_2$ are integers.

\begin{figure}[h]
\centerline{\hbox{
  \psfig{figure=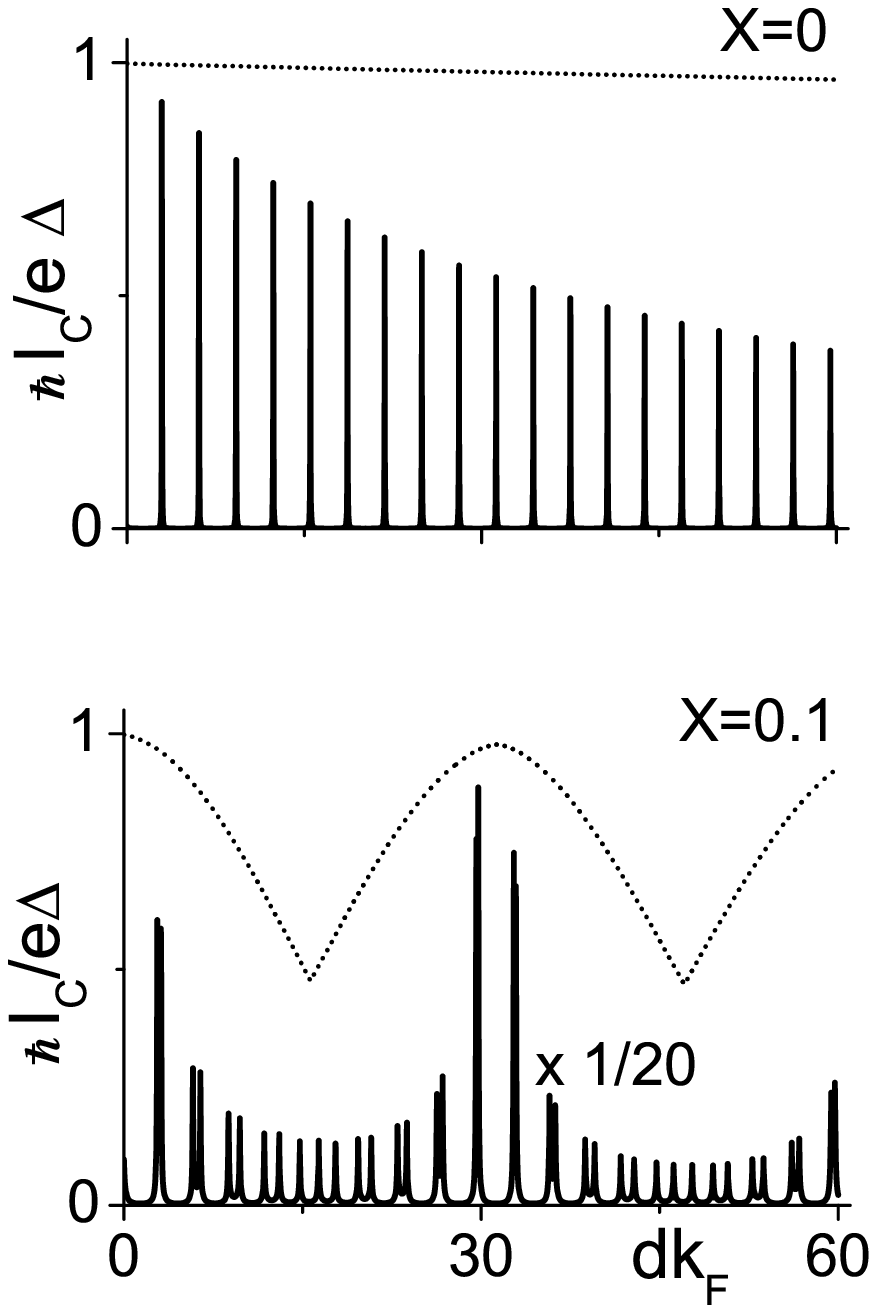,height=100mm,width=65mm,angle=0}
}}
  \caption{
   }
\label{PetkovicFig2}
\end{figure}

\begin{figure}[h]
\centerline{\hbox{
  \psfig{figure=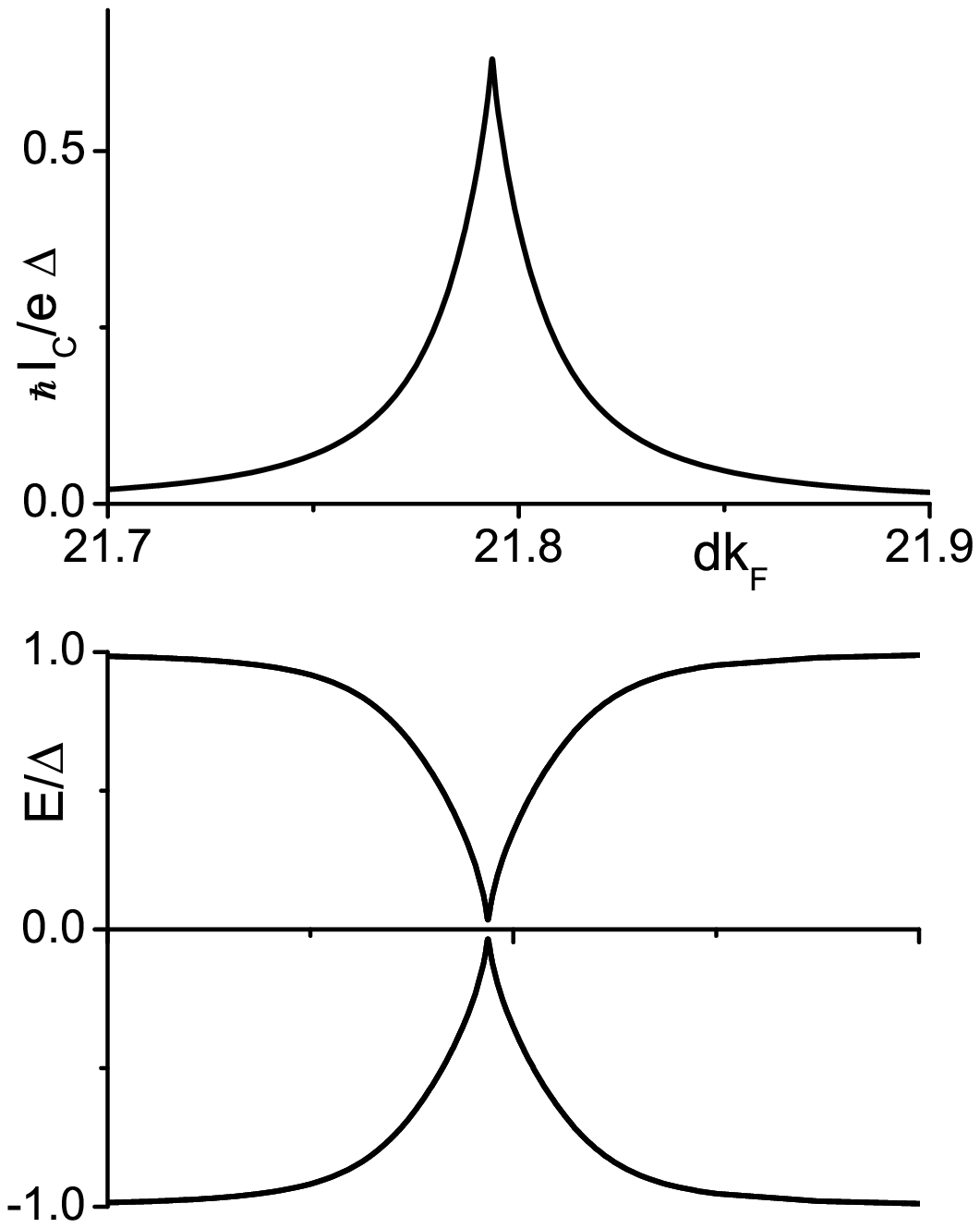,height=100mm,width=65mm,angle=0}
}}
  \caption{
  } \label{PetkovicFig3}
\end{figure}

The results are illustrated in
Figs.~\ref{PetkovicFig2}--\ref{PetkovicFig11}. In all illustrations
superconductors are characterized by $\Delta /{E_F}^{(S)} =
10^{-3}$, the Fermi energies of the two metals are equal,
${E_F}^{(F)} = {E_F}^{(S)}$, in F or N interlayers the normalized
exchange energy is $X=h/ {E_F}^{(F)} = 0.1$ or $0$, and the
insulating barriers are assumed to be equal and characterized by the
common parameter $Z_1=Z_2=Z$, related to the one-barrier
transmission probability via $|\hat{t}_{I_1}|^2 = |\hat{t}_{I_2}|^2
= 1/(1+Z^2)$.

In SINIS double-tunnel junctions sharp peaks occur in the critical
Josephson current as the junction width is varied
(Fig.~\ref{PetkovicFig2}, top panel) because of resonant
amplification of the Andreev process when the quasibound states of
the INI interlayer enter the superconducting gap and lower the
phase-sensitive Andreev bound states down to the Fermi level
(Fig.~\ref{PetkovicFig3}). The order of magnitude of the peaks is
the same as  the critical Josephson current values in transparent
junctions. Between two peaks, the junction behaves as a usual SIS
junction. In this case, quasibound states are not in the
superconducting gap.

In SIFIS junctions the critical current significantly decreases if
the junction transparency is reduced (Fig.~\ref{PetkovicFig2},
bottom panel). Instead of the critical current reaching a peak
value, in the case of low transparency a narrow dip opens up exactly
at the peak due to the onset of the $0$ -- $\pi$ transition. Note
that the modulation of spike heights in $I_C(d)$ is due to the
modulation of the Zeeman splitting. When the Zeeman-split quasibound
states are closer, the amplification is larger, the peaks tend to
the SNS value, and there is no triggering of the $0$ -- $\pi$
transition. Therefore, only the large enough spin splitting of sharp
enough quasibound states can trigger the $0$ -- $\pi$ transition. At
the transitions, spin-unpolarized Andreev states (with opposite spin
orientation on the same side of the Fermi surface) become spin
polarized, and vice versa. This is shown for two neighboring peaks
in Fig.~\ref{PetkovicFig4}. We point out that, in contrast to the
one-barrier problem,\cite{barash,Nikolai_SFSpi,nick_jetp} or
transparent interfaces, neither spin-polarized Andreev states have
to correspond to the $\pi$ state of the junction, nor
spin-unpolarized to the $0$ state. The following generalization is
valid: every $0$ - $\pi$ transition is accompanied by a switching of
the spin polarization of Andreev bound states,
Fig.~\ref{PetkovicFig5}. For low transparency, switching of the spin
polarization is not always accompanied by $0$ - $\pi$ transitions.
Partial contributions of Andreev states to $I(\phi)$ are shown in
Fig.~\ref{PetkovicFig6}. In the transition region, there is a
coexistence of stable and metastable $0$ and $\pi$ states due to a
considerable contribution of the second harmonic in $I(\phi)$,
Fig.~\ref{PetkovicFig6}.\cite{2001} In the tunnel limit,
$Z\to\infty$, the Andreev bound states that are practically $\phi$
independent coincide with bound states of the corresponding isolated
normal-metal interlayer, but with nonmatching spin polarization.
Because of this, the critical current is extremely weak.

With increased barrier transparency, the  mechanism of $0$ - $\pi$
transitions, described above, is modified by a broadening and
overlap of quasibound states in the normal interlayer. In the case
of complete transparency, the critical current modulation is related
to only two transitions in the considered interval of $dk_F$,
Fig.~\ref{PetkovicFig2}, bottom panel. This modulation is due to a
finite width of quasibound states even for transparent interfaces
because of the mismatch of Fermi energies for different spins
resulting in a finite normal reflection.\cite{Milos} Note that also
for finite, but relatively large transparency, $Z\sim 1$, there are
the same two $0$ - $\pi$ transitions as in fully transparent
junctions, although the critical current exhibits rapid oscillations
due to the resonant amplification. The same result holds for the
corresponding planar junctions after summation over transverse
channels, even for a small transparency of interfaces.\cite{greek}

Temperature-induced transitions both from $0$ to $\pi$ and from
$\pi$ to $0$ are studied by computing the phase diagram with
temperature and junction width as variables, for different
interfacial transparencies from metallic to tunnel limit,
Fig.~\ref{PetkovicFig7}. In the case of low transparency there is a
$0$ -- $\pi$ transition for each $d$ where the quasibound states
cross the Fermi surface.  The characteristic nonmonotonic dependence
of $I_C(T)$ with a well-pronounced dip at the transition is shown in
Figs.~\ref{PetkovicFig8} and \ref{PetkovicFig9}, with the
corresponding parts of the phase diagram in the insets. One can see
that the $\pi$ state can correspond to lower temperatures than the
$0$ state. In Figs.~\ref{PetkovicFig10} and ~\ref{PetkovicFig11}
Andreev levels and their contributions to $I(\phi)$ are shown for
three temperatures. Note that the discussed mechanism of $0$ - $\pi$
transition is more general and robust against the insertion of
scattering centers into the F interlayer. Similar results are
previously obtained for a diffusive point contact connecting two SF
bilayers (SFcFS).\cite{golubov_revmod}

\section{Conclusion}

We have shown the resonant nature of $0$ -- $\pi$ transitions in
short ballistic SIFIS junctions with one transverse channel open for
propagation. Unlike the corresponding SINIS structures, in
low-transparency SIFIS junctions the peaks in $I_C(d)$ curves do not
reach the corresponding value for transparent junctions; instead,
the $0$ - $\pi$ transitions are triggered and narrow dips open up
exactly at the peaks. This is not the case for higher interface
transparency, nor for planar junctions.

We have shown that the rule: "The $\pi$ state is characterized by
net spin polarization of the current-carrying Andreev
levels"\cite{tanaka,barash} is not quite general. Remaining true for
high-transparency junctions, in SIFIS tunnel junctions the situation
can be the opposite. We generalized the previous rule to read as
follows: $0$ - $\pi$ transition is accompanied by the switching of
net spin polarization of current-carrying Andreev levels. In
agreement with experimental observations,\cite{ryazanov} we also
demonstrated that in the phase diagram of double-barrier SIFIS
junctions the $\pi$ state can occur at lower temperatures than the
$0$ state (or the other way round), depending on parameters. This
also holds for planar double-barrier junctions,\cite{greek} as well
as for diffusive SFcFS point contacts.\cite{golubov_revmod}

\bigskip

\section*{ACKNOWLEDGMENTS}

This work has been supported by the Serbian Ministry of Science,
Project No.~141014, and by Franco-Serbian PAI EGIDE Project
No.~11049XG. NMC also acknowledges the support of RFBR Project
No.~03-02-16677, No.~04-02-08159, and No.~06-02-17519, the Russian
Ministry of Science, the Netherlands Organization for Scientific
Research NWO, CRDF, and Russian Science Support Foundation.

\vspace{4mm} Fig. 1 A schematic of a double-barrier
superconductor-ferromagnet-superconductor junction. $I_1$ and $I_2$
denote the insulating layers.

Fig. 2 Critical Josephson current $I_C$ as a function of the
normal-metal-interlayer thickness $d$ at zero temperature for
transparent ($Z=0$, dotted curves) and low-transparency ($Z = 10$,
solid curves) interfaces, for the SINIS junction ($X=0$, top panel)
and the SIFIS junction ($X = 0.1$, bottom panel). Note that $I_C(d)$
for low-transparency SIFIS junction is magnified $20$ times.

Fig. 3 A peak in $I_C(d)$ and the spin degenerate Andreev levels,
$E(\phi_C)$, for a SINIS junction with $Z=10$ and $X = 0$ at zero
temperature.

Fig. 4 Two neighboring peaks in $I_C(d)$ for the SIFIS junction with
$X = 0.1$ and $Z=10$, and spin split Andreev levels, $E(\phi_C)$:
two spin up levels (thick and thin solid curves) and two spin down
levels (thick and thin dashed curves). Insets: dips at
$0\leftrightarrow \pi$ transitions.

\begin{figure}[h]
\centerline{\hbox{
  \psfig{figure=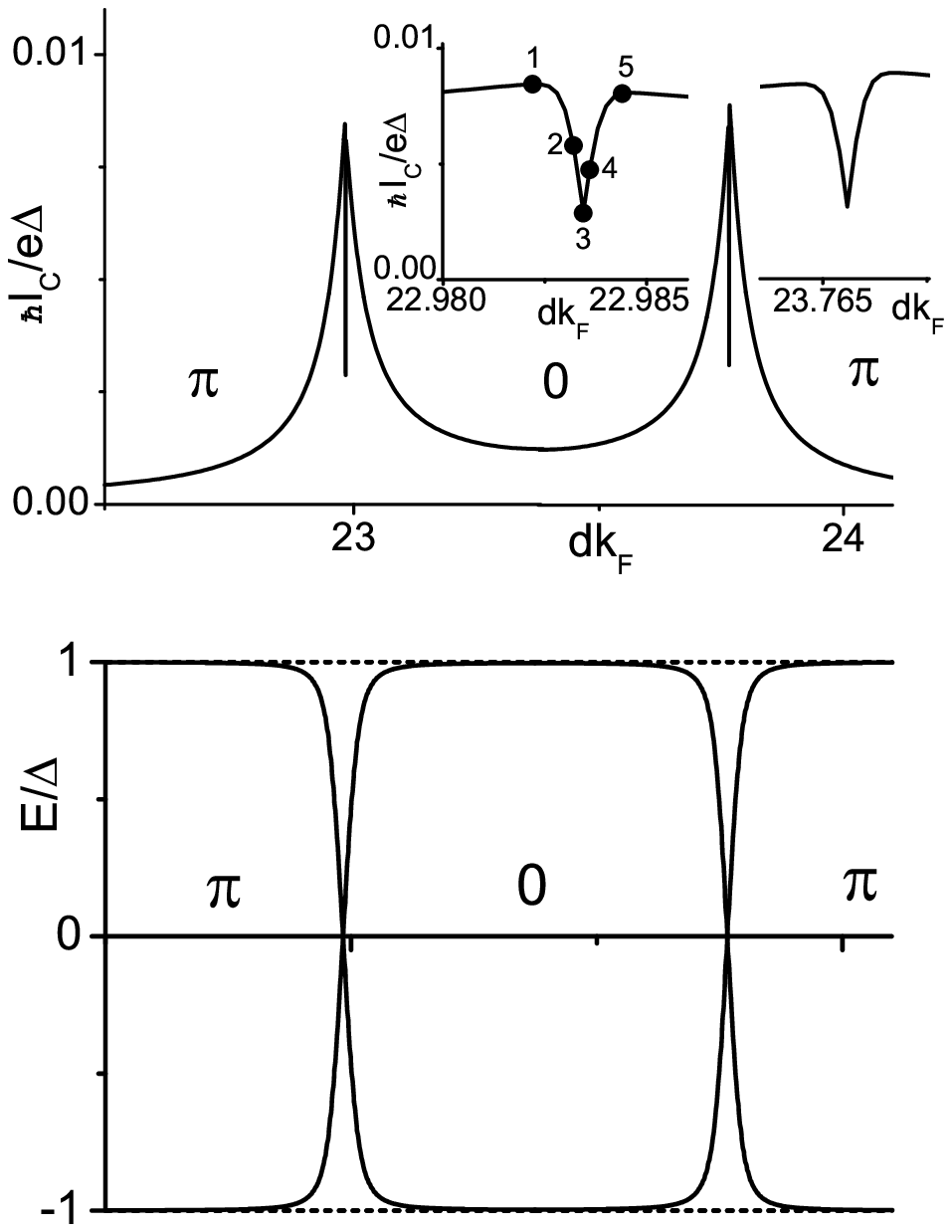,height=160mm,width=100mm,angle=0}
}}
  \caption{
  } \label{PetkovicFig4}
\end{figure}

Fig. 5 Variation of the Andreev levels as a function of the phase
difference $\phi$
  for five values of the normal-layer thickness indicated in the
  inset of Fig.~\ref{PetkovicFig4}, with all other parameters
  being the same. The arrows indicate the spin polarization of the
  levels.

\begin{figure}[h]
\centerline{\hbox{
  \psfig{figure=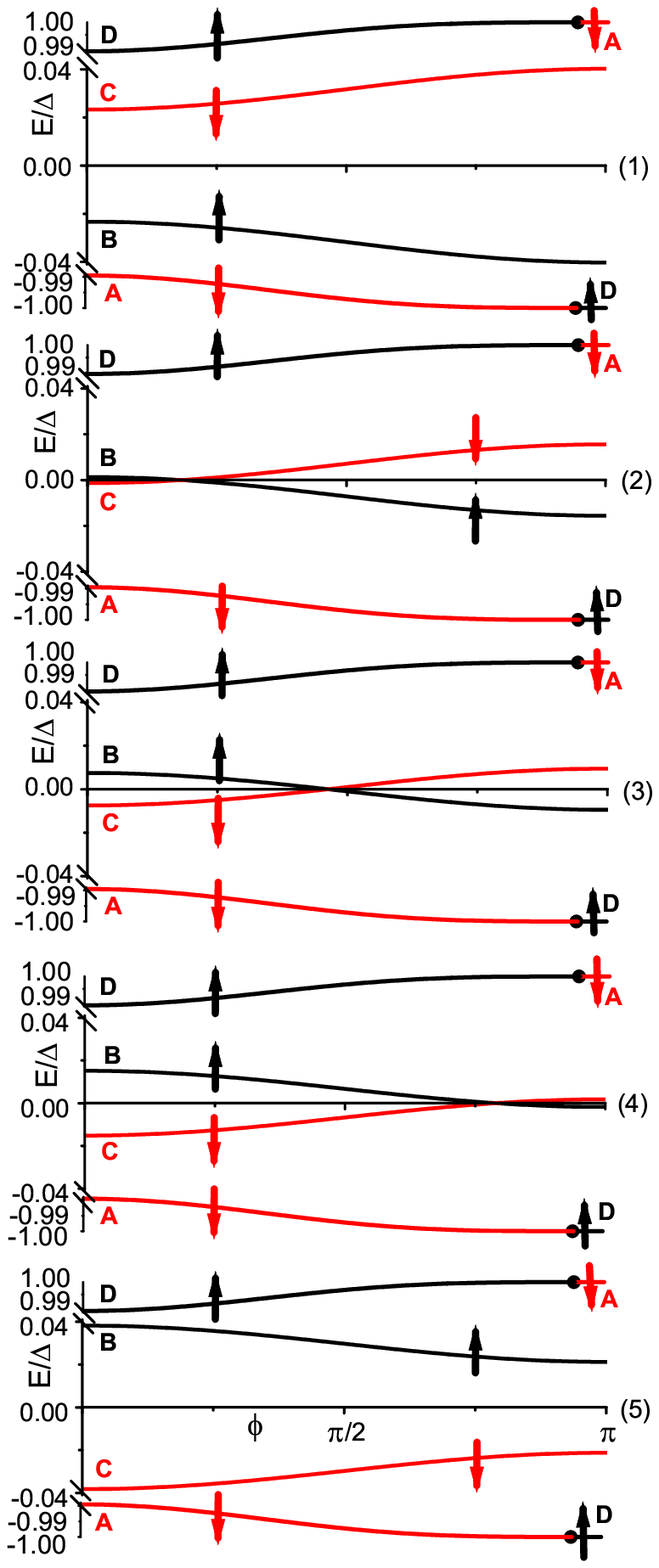,height=180mm,width=65mm,angle=0}
}}
  \caption{
  } \label{PetkovicFig5}
\end{figure}

\begin{figure}[h]
\centerline{\hbox{
  \psfig{figure=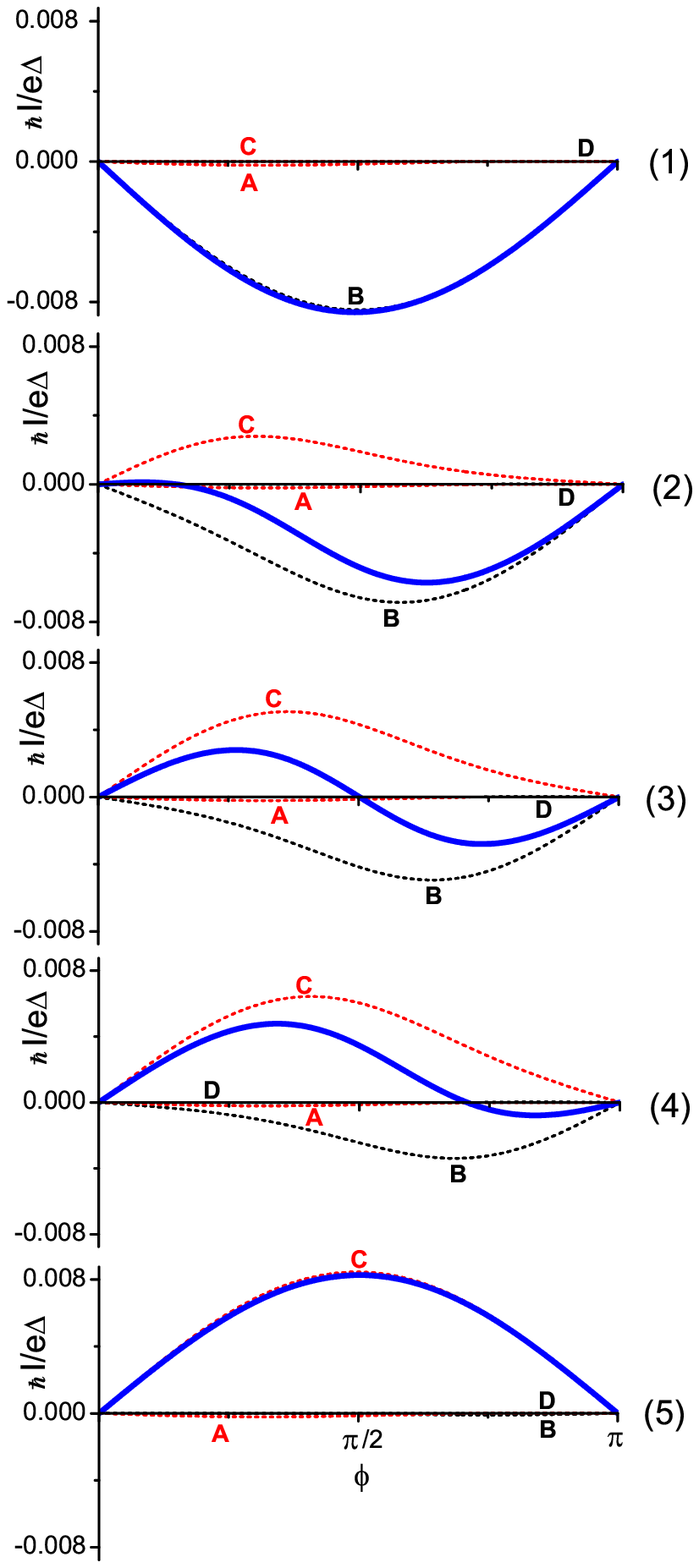,height=180mm,width=65mm,angle=0}
}}
  \caption{
   } \label{PetkovicFig6}
\end{figure}

\begin{figure}[h]
\centerline{\hbox{
  \psfig{figure=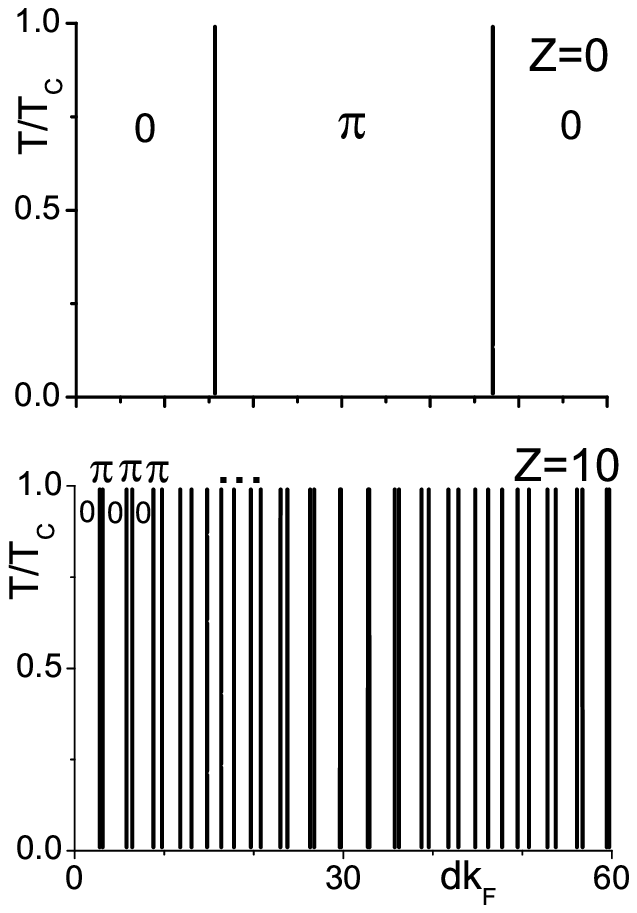,height=100mm,width=75mm,angle=0}
}}
  \caption{
  } \label{PetkovicFig7}
\end{figure}

\begin{figure}[h]
\centerline{\hbox{
  \psfig{figure=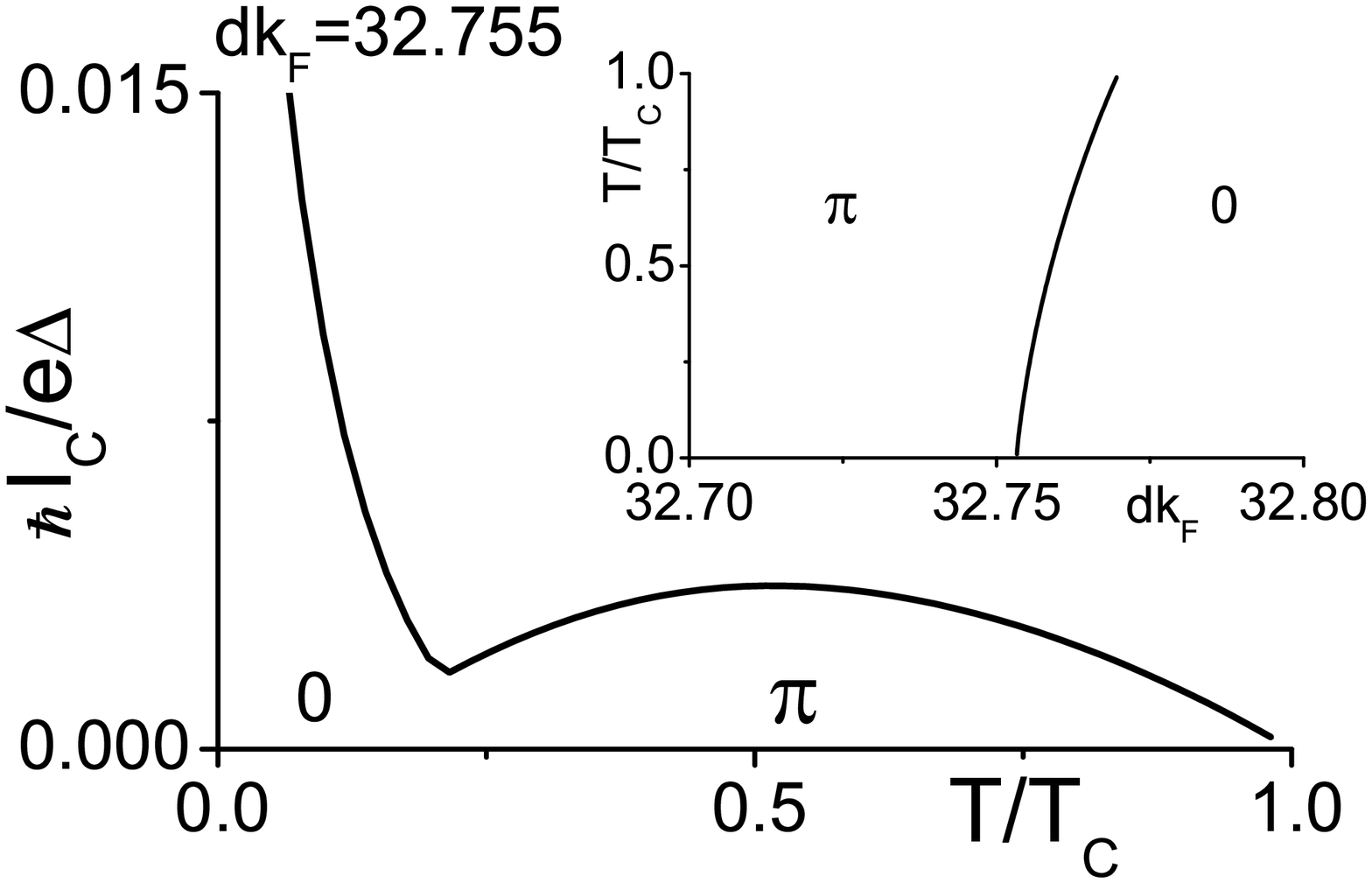,height=65mm,width=75mm,angle=0}
}}
  \caption{
   } \label{PetkovicFig8}
\end{figure}

\begin{figure}[h]
\centerline{\hbox{
  \psfig{figure=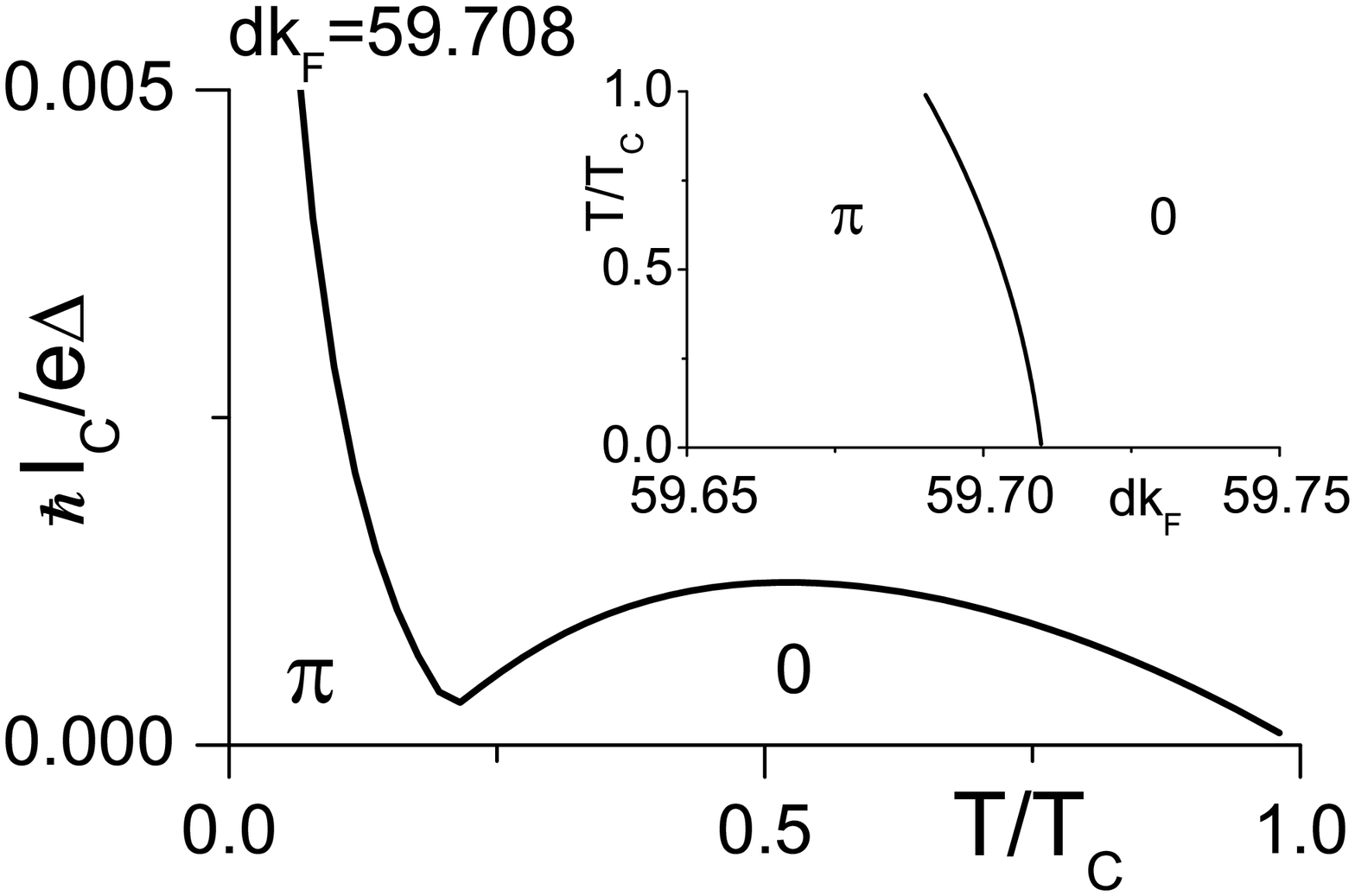,height=65mm,width=75mm,angle=0}
}}
  \caption{
  } \label{PetkovicFig9}
\end{figure}

Fig. 6 The current-phase relation (solid curves) for
  five values of the normal-layer thickness indicated in the inset
  of Fig.~\ref{PetkovicFig4}, with all other parameters being the same. Partial currents carried by the corresponding Andreev
  levels shown in Fig.~\ref{PetkovicFig5} are also presented (dotted curves).

Fig. 7 The phase diagram as a function of the normalized temperature
$T/T_C$ and normalized ferromagnetic-layer thickness $dk_F$, for
transparent ($Z=0$, top panel) and low-transparency interfaces
($Z=10$, bottom panel), and for $X=0.1$.

Fig. 8 Characteristic temperature variation of $I_C$ showing the
  temperature-induced transition from $0$ to $\pi$ states for a
junction width $dk_F=32.755$, $X=0.1$, and $Z=10$. The inset
  shows the part of the phase diagram for the corresponding transition.

Fig. 9 Characteristic temperature variation of $I_C$ showing the
  temperature-induced transition from $\pi$ to $0$ states, for a junction width
  $dk_F=59.708$, $X=0.1$, and $Z=10$. The inset
  shows the parts of the phase diagram for the corresponding transition.

Fig. 10 Variation of the Andreev levels as a function of the phase
difference $\phi$ for $dk_F=29.609$, $X=0.1$, and $Z=10$. The arrows
indicate the spin polarization of the levels.

\begin{figure}[h]
\centerline{\hbox{
  \psfig{figure=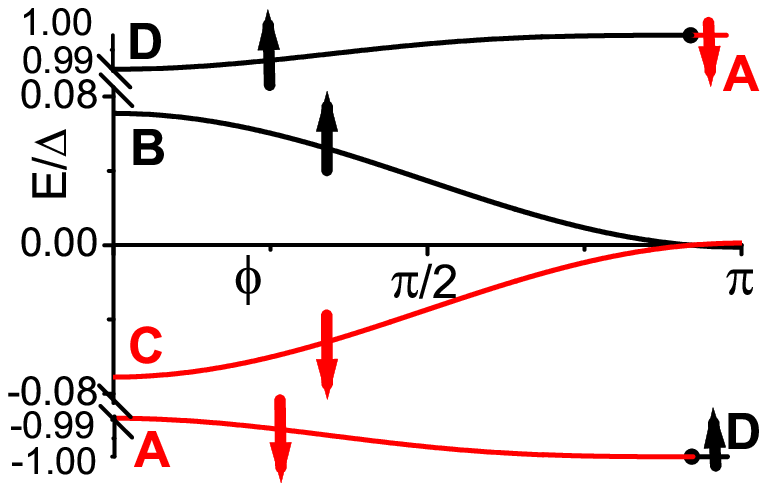,height=50mm,width=60mm,angle=0}
}}
  \caption{
} \label{PetkovicFig10}
\end{figure}

\begin{figure}[h]
\centerline{\hbox{
  \psfig{figure=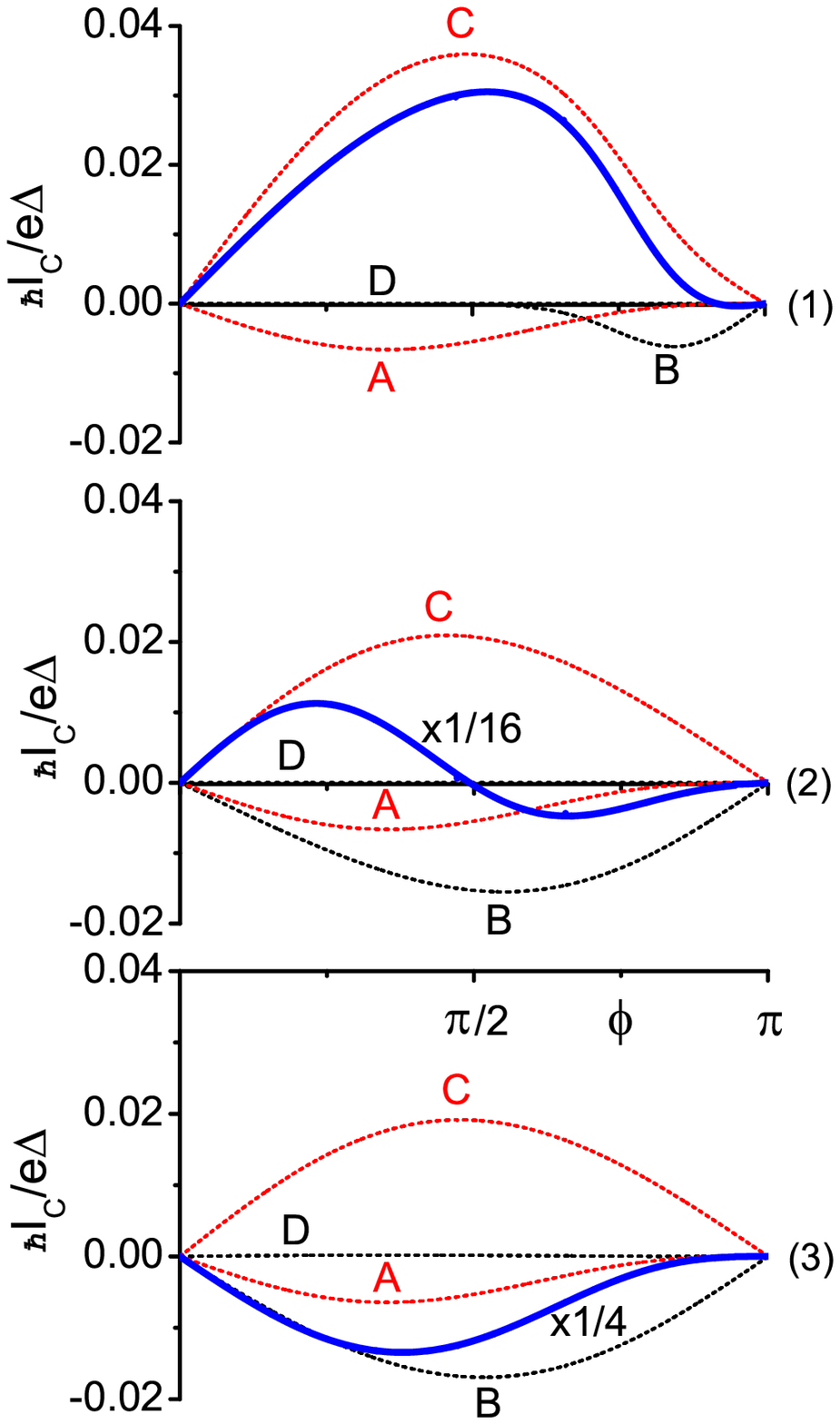,height=130mm,width=65mm,angle=0}
}}
  \caption{
   } \label{PetkovicFig11}
\end{figure}

Fig. 11 The current-phase relation (solid curves) for the same
parameters as in the Fig.~\ref{PetkovicFig10}, and for three
different temperatures: ($1$)~$T/T_C=0.01$,~($2$)~$T/T_C=0.2$,
and~($3$)~$T/T_C=0.5$. Partial currents carried by the corresponding
Andreev levels are also presented (dotted curves).

\bigskip
\newpage

\end{document}